\documentclass{aa}
\usepackage{threeparttable}
\usepackage{natbib}
\usepackage{xcolor}
\usepackage{placeins}

\usepackage{microtype}
\usepackage{graphicx}
\usepackage{txfonts}
\PassOptionsToPackage{hyphens}{url}
\usepackage{hyperref}
\defcitealias{Leaman2021}{LvdV21}

\begin{document}

\title{Disentangling the formation mechanisms of nuclear star clusters}
 \author{Katja Fahrion\inst{1,2}
          \and
          Ryan Leaman\inst{3,4}
          \and
          Mariya Lyubenova\inst{1}
          \and
          Glenn van de Ven\inst{4}
          }

   \institute{European Southern Observatory, Karl Schwarzschild Stra\ss{}e 2, 85748 Garching bei M\"unchen, Germany
        \and
        European Space Agency, European Space Research and Technology Centre, Keplerlaan 1, 2200 AG Noordwijk, Netherlands\\
        \email{katja.fahrion@esa.int}
        \and
   Max-Planck-Institut f\"ur Astronomie, K\"onigstuhl 17, 69117 Heidelberg, Germany
   \and
   Department of Astrophysics, University of Vienna, T\"urkenschanzstrasse 17, 1180 Wien, Austria}
   \date{}

 \abstract{Nuclear star clusters (NSCs) are massive star clusters found ubiquitously in the centres of galaxies, from the dwarf regime to massive ellipticals and spirals. The fraction of nucleated galaxies is as high as $>$ 90 \% at $M_{\text{gal}} \sim 10^9 M_\sun$. However, how NSC formation mechanisms work in different regimes and what determines galaxy nucleation is still unclear.  The dissipationless accretion of infalling globular clusters (GCs) and the in situ formation of stars directly at the galactic centre likely operate to grow NSCs in most galaxies; however, their efficiency has been difficult to assess observationally.
 Here, we provide, for the first time, a quantitative determination of the relative strength of these processes in the build-up of individual NSCs. Using a semi-analytical model of NSC formation based on the orbital evolution of inspiraling GCs, together with observed NSC and GC system properties, we derived the mass fraction of in situ born stars $f_\text{in, NSC}$ for 119 galaxies with masses from $3 \times 10^{7}$ to $3 \times 10^{11} M_\odot$, in the Local Volume, the Fornax, and Virgo galaxy clusters. Our analysis reveals that the NSC mass, as well as the ratio of NSC to the total GC system mass, are strong indicators of the dominant NSC formation channel, and not the total galaxy stellar mass as previously suggested.
 More massive NSCs formed predominantly via the in situ formation of stars ($f_\text{in, NSC} \sim 0.9$), while the lower-mass NSCs are expected to have formed predominantly through the merger of GCs ($f_\text{in, NSC} \sim 0.2$). The results of this simple model are in agreement with recent independent estimates of the dominant NSC formation channel from recent stellar population analysis.}
   \maketitle

\section{Introduction}
\label{sect:intro}
Nuclear star clusters (NSCs) are dense, massive star clusters located at the centre of galaxies. Typical NSCs have masses and sizes between $10^5 - 10^8 M_\sun$ and $3 - 10$ pc, respectively \citep{Boeker2004, Walcher2005, Turner2012, Cote2006, denBrok2014, Georgiev2016, Spengler2017}.
The nucleation fraction of galaxies was recently found to reach up to $>$ 90 \% at stellar masses of $M_\ast \sim 10^9 M_\sun$, with a drop down to $\sim$ 10 \% for both the lowest and highest galaxy masses (10$^6$ and $10^{11} M_\sun$, respectively, \citealt{SanchezJanssen2019, Carlsten2021, Hoyer2021}).

Although NSCs are common, their formation mechanisms are still under debate. Two pathways are typically discussed (see the review \citealt{Neumayer2020}): the (gas-free) accretion of globular clusters (GCs) that spiral inwards due to dynamical friction \citep{Tremaine1975, CapuzzoDolcetta1993, CapuzzoDolcetta2008, Agarwal2011, ArcaSedda2014}, and an in situ formation scenario, where the NSC forms at the galaxy centre from infalling gas and local star formation \citep{Loose1982, Milosavljevic2004, Schinnerer2007, Bekki2006, Bekki2007, Antonini2015}. In addition, work by \cite{Guillard2016} has shown that these processes may operate together in a hybrid scenario of gas-rich proto-star cluster accretion.

Detailed $N$-body simulations of the dynamical evolution of inspiralling GCs have successfully reproduced the structural properties of nearby NSCs such as their density profiles, ellipticity, and angular momentum \citep{Hartmann2011, Antonini2012, Tsatsi2017, Lyubenova2019}. Semi-analytical models generalising the evolution of GC systems (GCSs) have reproduced some average trends for NSC-host scaling relations \citep{Antonini2012, Antonini2013, Gnedin2014}.  However, the intrinsic scatter in these relations, as well as the shape of the  NSC formation efficiency curve with galaxy mass, has been difficult to address given the number of physical processes and scales at play. Nevertheless, the GC-accretion scenario provides a natural explanation for the presence of metal-poor stellar populations observed in some NSCs \citep{Rich2017, AlfaroCuello2019, Johnston2020, Fahrion2020a, Fahrion2021}.

The in situ formation channel is usually invoked to explain the presence of young, metal-rich stellar populations found in many NSCs hosted by late-type galaxies (LTG) \citep{FeldmeierKrause2015, Carson2015, Kacharov2018, Nguyen2017, Nguyen2018}. The details of the gravitational potential and resonances in individual galaxies are key in driving gas flows to central regions \citep{Sormani18}, while the complex radiation field in galactic nuclei prevents a simple understanding of how efficiently star formation proceeds \citep{Longmore2018}.

To explain the full spectrum of structural, kinematic, and stellar population properties of individual NSCs, both processes are most likely required \citep{Hartmann2011, Lyubenova2013, Antonini2013, Antonini2015, Guillard2016}. Recently, \cite{Neumayer2020} suggested a transition of the dominant NSC formation channel from GC accretion to in situ formation at galaxy masses of $M_\text{gal} \sim 10^9 M_\sun$ based on the functional shape of the nucleation fraction of galaxies and stellar population properties of the NSCs of low-mass early-type galaxies (ETGs, \citealt{Paudel2011, Spengler2017}) that tend to be more metal-poor than the expected metallicity of the host as suggested by the mass-metallicity relation \citep{Gallazzi2005, Kirby2013}. Similar trends were found in dwarf galaxies, where both the metallicity of NSCs and hosts were measured \citep{Leaman2020, Fahrion2020a, Johnston2020}.

Recently, \cite{Fahrion2021} presented a detailed stellar population analysis of 25 nucleated ETGs. Based on metallicity profiles and the background-cleaned star formation histories of the NSCs in comparison to the host galaxy, the dominant NSC formation channel was identified in the individual galaxies. This work shows that low-mass NSCs in low-mass galaxies ($M_\text{NSC} < 10^7 M_\sun$ and $M_\text{gal} < 10^{9} M_\sun$, respectively) are predominantly formed from the inspiral of a few GCs as evident from metal-poor populations in the NSCs. More massive NSCs in more massive galaxies likely build most of their mass through central star formation, giving rise to extended star formation histories, high metallicities, and sometimes young stellar populations. At intermediate masses ($M_\text{gal} \sim 10^9 M_\sun$), the analysis presented in \cite{Fahrion2021} identified signatures of both channels acting in the same galaxy.

Given these recent results, we aim to build a simple model for uncovering NSC formation pathways that allows us to understand the trends with galaxy and NSC mass. Detailed stellar population analysis is currently restricted to a handful of galaxy-NSC systems, but many more observations are available based on photometry (e.g. \citealt{Turner2012, Ordenes2018, SanchezJanssen2019, Carlsten2021}). Therefore, we aim to test a NSC formation model using photometric measurements of a large sample of galaxies. Such a model should be applicable to individual host galaxies and be flexible enough to predict properties of the NSC and the galaxy's surviving star cluster population.
For systems with similar masses or star formation histories, one may expect correlated differences between the NSC properties and the statistics of the surviving GC populations if their NSCs formed in purely in situ or purely merger-based scenarios.

In this work, we use a new semi-analytical model of NSC formation to derive the mass fraction of individual NSCs that formed via in situ star formation ($f_\text{in, NSC}$ = $1 - f_\text{acc, NSC}$) in galaxies with stellar masses $3 \times 10^{7}$ to $3 \times 10^{11} M_\odot$. Section\,\ref{sect:model} introduces the relevant parts of the model and Sect.\,\ref{sect:sample} describes the sample of nucleated galaxies to which we applied this.  The key results are shown in Sect. \ref{sect:results} and discussed in Sect.\,\ref{sect:discussion}. We conclude in Sect.\,\ref{sect:conclusion}.

\section{The model}
\label{sect:model}

We used an analytical model of NSC formation that was presented in \citealt{Leaman2021} (hereafter \citetalias{Leaman2021}). While other studies have attempted to model a large variety of complex physical processes involved in the mass build-up of NSCs, this analytic model describes NSC formation solely via the accretion of GCs into the centre due to dynamical friction on a statistical basis. It explores how many of the total GC and NSC population characteristics can be reproduced when only considering the GC inspiral scenario. The model considers simple analytic prescriptions for dynamical friction and the GC mass distributions to describe the orbital evolution of GCs that sink to the galaxy centre and simultaneously lose mass in the tidal field.  Key to this model is the dependence of the host galaxy baryonic structure (size and mass) in modulating the efficiency of NSC formation, and the mass budget of the NSCs and surviving GCs.

\subsection{Limiting star cluster mass for inspiral}
The basis for the predictions presented in \citetalias{Leaman2021} is to consider the limiting mass for a GC that is needed for inspiral to the centre without being disrupted in the tidal field. Their work identified this limiting mass $M_\text{GC, lim}$ by considering dynamical friction-driven orbital inspiral from an initial starting position in the galaxy for a variety of GC mass functions and host galaxy parameters.

This limiting mass depends on the galaxy mass and size, but roughly scales as $M_\text{GC, lim} \propto R_\text{gal}^{2}$. GCs more massive than this limit are expected to have sunk to the centre and contribute to the mass build-up of the NSC, while less massive GCs might be still observed in the galaxy today. Importantly, \citetalias{Leaman2021} show that the theoretically predicted limiting masses agreed very well with the observed most massive GC in the same galaxies we consider here. As described later, in the application of the model presented here, we typically use the observational estimates on $M_\text{GC, lim}$ from the most massive GC observed in each system.

\subsection{NSC and GC system masses}
From a total population of GCs, the model predicts a host galaxy-dependent limiting mass $M_\text{GC, lim}$, above which GCs contribute to the NSC.
An important aspect of the model is that it predicts not only the nucleation probability, but also gives limits and expectation values of the mass of the NSC and the mass of the surviving GC populations in individual galaxies.

We used the following description of the total mass in star clusters\footnote{We note that we use star cluster and GC interchangeably.} from \cite{Elmegreen2018}:
\begin{equation}
    M_\text{cl, total} = M_\text{cl, max}\left[1 + \text{ln}\left(\frac{M_\text{cl, max}}{M_\text{cl, min}}\right)\right],
    \label{eq:cluster_total_mass}
\end{equation}
where $M_\text{cl, max}$ and $M_\text{cl, min}$ are the most massive and least massive star clusters that have ever formed. This equation is based on integrating an initial power-law mass function of star clusters with a slope of $\beta = -2$.
To avoid a divergence in the integral, the assumption is made that the mass function includes a single cluster with mass $M_\text{cl, max}$, which does not have the same meaning as a theoretical truncation mass\footnote{For simplicity, we chose to follow the description in \cite{Elmegreen2018} and do not to include a theoretical truncation as an additional parameter as this is not constrained. Testing with such an additional parameter has shown that the inferred masses change by less than a factor of 1.5 and the inferred in situ mass fractions do not significantly differ from the ones derived in our default approach.}.

This total mass in star clusters is some fraction $\eta$ of the total galaxy mass $M_\text{gal}$:
\begin{equation}
    M_\text{cl, total} = \eta M_\text{gal}.
\end{equation}
GCs more massive than the limiting mass $M_\text{GC, lim}$ spiral to the centre to form the NSC. Therefore, using the same argument as for Eq. \ref{eq:cluster_total_mass}, the expected NSC mass from GC accretion is given as follows:
\begin{equation}
M_\text{NSC, acc} = M_\text{cl, max} \left[1 + \text{ln}\left(\frac{M_\text{cl, max}}{M_\text{GC, lim}}\right)\right].
\label{eq:M_NSC_initial}
\end{equation}
Combining the previous equations, the NSC mass from accreted GCs is given as follows:
\begin{equation}
    M_\text{NSC, acc} = \eta\,M_\text{gal} \left[\frac{1 + \text{ln}\left(\frac{M_\text{cl, max}}{M_\text{GC, lim}}\right)}{1 + \text{ln}\left(\frac{M_\text{cl, max}}{M_\text{cl, min}}\right)} \right].
    \label{eq:NSC_acc}
\end{equation}
For GCs less massive than the model threshold $M_\text{GC, lim}$, a subset with masses below a dissolution mass $M_\text{diss}$ is destroyed in the tidal field, while the remainder survives as the observed GCS. Therefore, the total mass of the GCS as observed today is expressed as follows:
\begin{equation}
    M_\text{GCS} = \eta\,M_\text{gal} - M_\text{NSC, acc} - M_\text{diss} \left[1 + \text{ln}\left(\frac{M_\text{diss}}{M_\text{cl, min}}\right)\right].
    \label{eq:M_GCS}
\end{equation}
We used these simultaneous predictions for the NSC and GCS masses to estimate the deficit in the NSC mass predicted only from GC accretion. For fixed host galaxy parameters, the fraction of the NSC mass assembled via in situ formation $f_\text{in, NSC}$ can, therefore, be estimated by scaling the predicted NSC mass ($M_\text{NSC, acc}$) to the observed mass of individual NSCs:
\begin{equation}
    M_\text{NSC} = (1 - f_\text{in, NSC})^{-1} M_\text{NSC, acc}.
\end{equation}
The model requires, as input, an estimate of the threshold mass for GC inspiral ($M_\text{GC, lim}$) which can be predicted from the size and mass of the host galaxy in the \citetalias{Leaman2021} model, or estimated based on the observed most massive surviving GC. \citetalias{Leaman2021} show that both methods agree within 0.26 dex, provided contaminating ultra-compact dwarf galaxies (UCDs) are removed from the GC field population. The free parameters in the model are a normalisation factor for the integrated fraction of stars that ever formed in clusters ($\eta$), as well as estimates of the most massive, least massive, and typical dissolution mass for GCs. To estimate $f_\text{in, NSC}$, the model predictions for $M_\text{NSC}$ and $M_\text{GCS}$ can be compared to those observed values in an individual galaxy.

We estimated the parameters for this model using \textsc{emcee} \citep{emcee}, a python implementation of the Markov Chain Monte Carlo (MCMC) method that samples the posterior probability distribution function. We note that $M_\text{NSC}$ and $M_\text{GCS}$ are input variables that are fitted by the model and the remaining parameters are introduced as priors. For $M_\text{gal}$ and $M_\text{diss}$, we used Gaussian priors given by the observed stellar mass and the observed least massive, surviving GC ($M_\text{GC, min}$) of an individual galaxy. The latter provides a crude observational estimate of the GC dissolution mass.

It is important to note that $M_\text{GC, lim}$ can be derived from the \citetalias{Leaman2021} model based on the galaxy mass and size - scaling as roughly $M_\text{GC, lim} \propto R_\text{gal}^{2}$. Alternatively, the most massive GC observed today ($M_\text{GC, max}$), should give an observational limit to $M_\text{GC, lim}$. Therefore, for galaxies with catalogues of individual GCs, we used $M_\text{GC, max}$ as an estimate for $M_\text{GC, lim}$. For other galaxies, we used the theoretical prediction from the \citetalias{Leaman2021} model. Possible effects relating to this choice of $M_\text{GC, lim}$ are discussed in Sect. \ref{sect:model_setup}. For $M_\text{GC, lim}$, we also used Gaussian priors centred on either the model prediction or the estimate based on the observed value of $M_\text{GC, max}$.
For the other parameters, we adopted flat priors as listed in Table \ref{tab:priors}, and $M_\text{cl, max}$ further has to be less than $\eta M_\text{gal}$. For the Gaussian priors on the galaxy and various star cluster masses, we adopted conservative uncertainties of 0.3 dex \citep{Mendel2014}.

\begin{table*}[h]
    \centering

    \caption{Priors used in the modelling.}
    \begin{tabular}{c|l c}
      Parameter   & Description & Prior  \\ \hline \hline
    $f_\text{in, NSC}$ & in situ NSC mass fraction & (0.0, 1.0) \\
    $\eta$ & mass fraction in clusters   & ($10^{-5}$, 0.5) \\
    $M_\text{gal}$ & host stellar mass & measured \\
    $M_\text{cl, min}$ & least massive GC ever & (10, 100) \\
    $M_\text{cl, max}$ & most massive GC ever & ($M_\text{GC, max}$, 7.6 $\times10^7$) and $< \eta M_\text{gal}$ \\
    $M_\text{GC, lim}$ & limiting mass for inspiral & $\sim M_\text{GC, max}$ or model prediction from \citetalias{Leaman2021} \\
    $M_\text{diss}$ & dissolution mass limit & $\sim$ $M_\text{GC, min}$ \\ \hline
    \end{tabular}
    \label{tab:priors}
    \tablefoot{Masses in $M_\sun$. Except for $f_\text{in, NSC}$, the priors are implemented in logarithmic units. Theoretical limit on $M_\text{cl, max}$ from \cite{Norris2019}.}
\end{table*}

\begin{figure}
    \centering
    \includegraphics[width=0.49\textwidth]{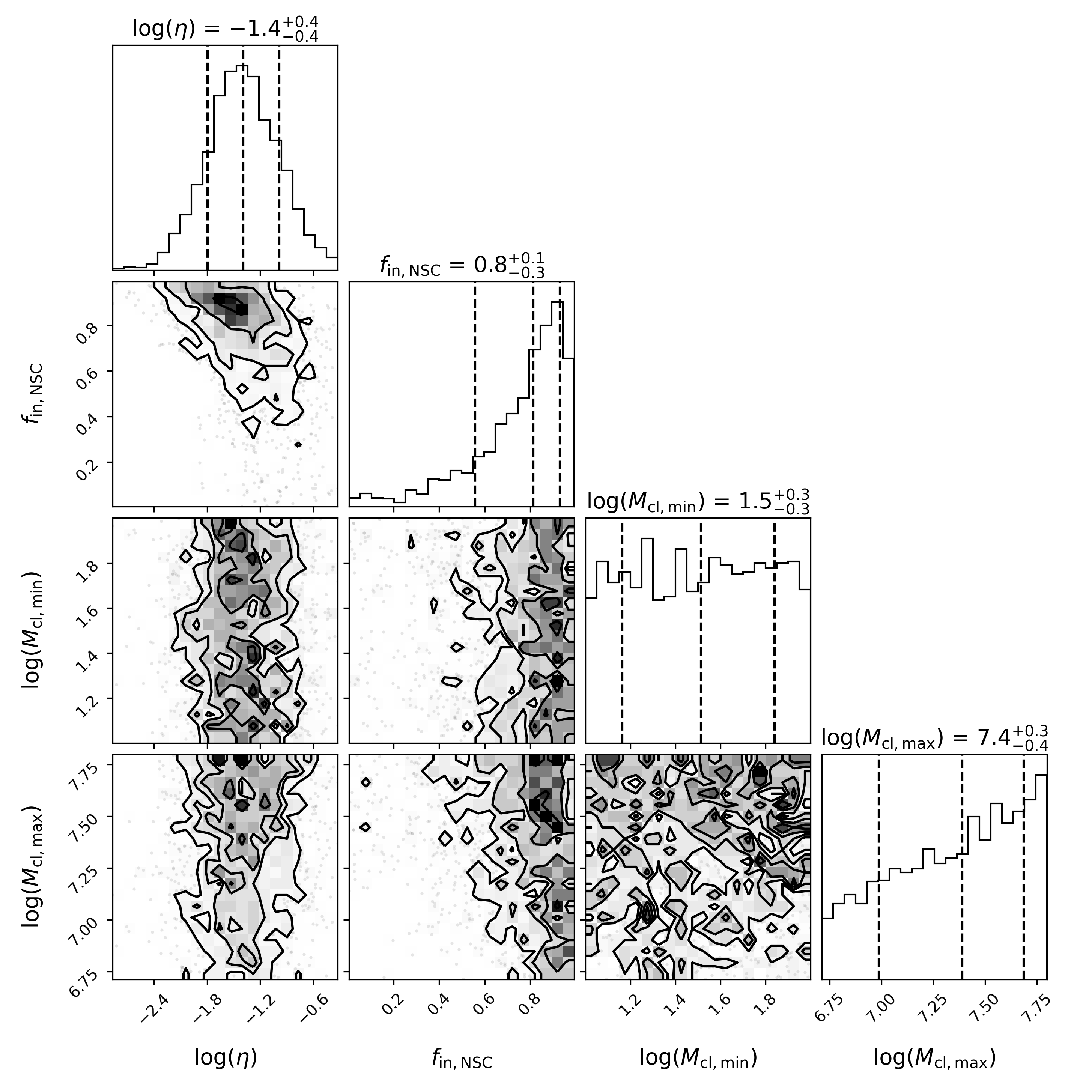}
    \caption{Posterior distribution for the MCMC fit of FCC\,47. We only show the parameters with flat priors. The other parameters follow Gaussian priors.}
    \label{fig:FCC47_corner}
\end{figure}

Figure \ref{fig:FCC47_corner} shows an example resulting posterior distributions for FCC\,47, an ETG in the Fornax cluster with a massive NSC ($M_\text{NSC} \sim 7 \times 10^{8} M_\sun$). Parameters with Gaussian priors are not shown for simplicity. For FCC\,47, the model finds a high $f_\text{in, NSC} = 0.8$, which is in agreement with the results from kinematic and stellar population analysis \citep{Fahrion2019b}.

We note that our model is only applicable for galaxies with both NSCs and GCs. In nature, galaxies can be non-nucleated, but still have a GCS. On the other hand, a few dwarf galaxies are known to only posses a single star cluster in their photometric centre (e.g. KKs58 and DDO216; \citealt{Fahrion2020a, Leaman2020}).
Although there are various processes that can prohibit NSC formation or destroy NSCs, these are not included in our model and we refer readers to \citetalias{Leaman2021} for a detailed discussion on how these relate to the nucleation fraction of galaxies.

\section{Data}
\label{sect:sample}
\begin{figure*}
    \centering
    \includegraphics[width=0.98\textwidth]{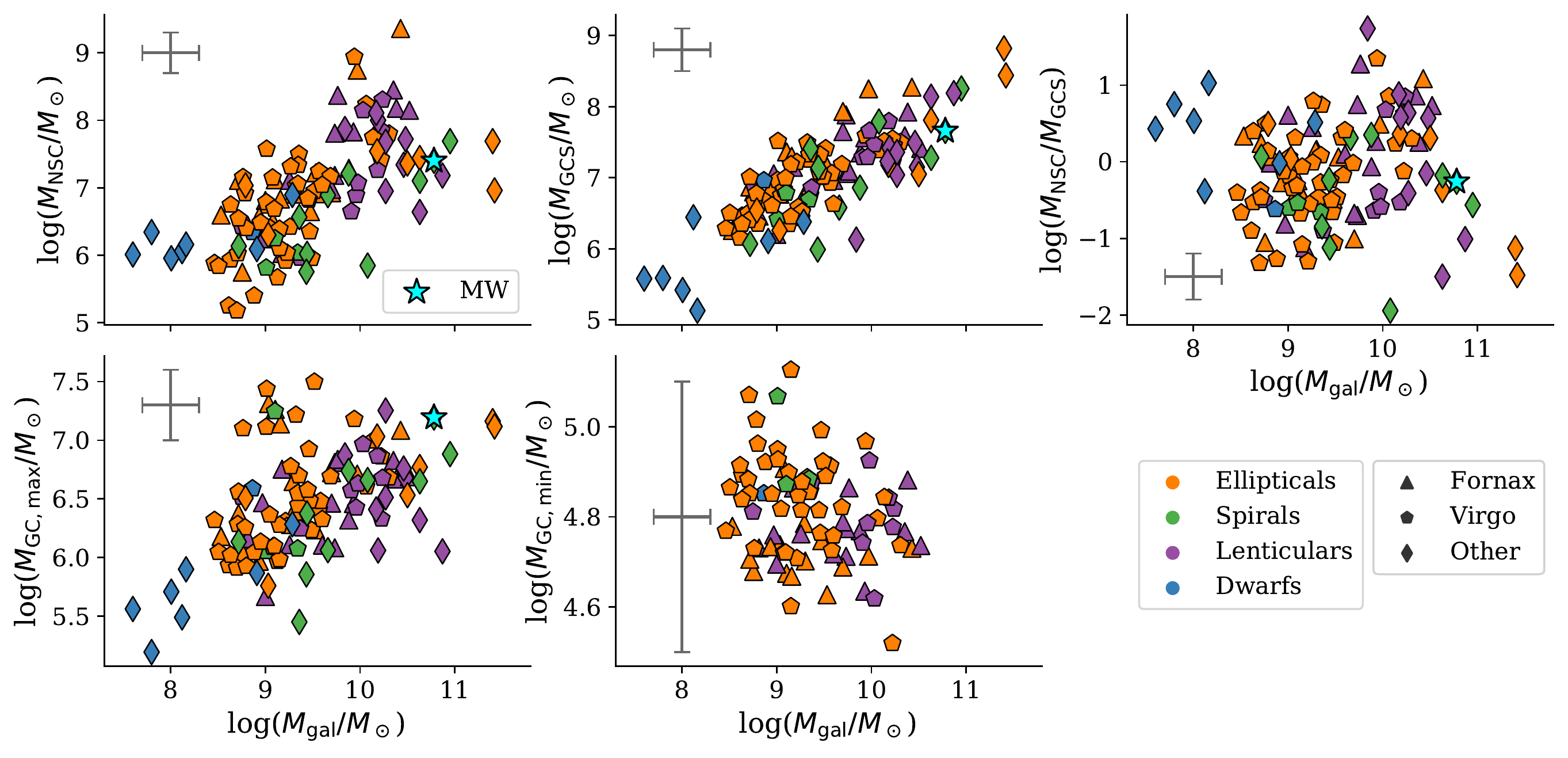}
    \caption{Sample of nucleated galaxies used in this study.
    The symbol types differentiate between galaxies in Fornax (triangle), Virgo (hexagon), and other environments (diamond), whereas the colours show the different galaxy types: ellipticals (orange), spirals (green), lenticulars (S0 and SB0, purple), and dwarf irregulars and spheroidals (blue). The Fornax and Virgo sample only contains ETGs and lenticular galaxies. The NSC of the Milky Way (MW) is shown in cyan. The grey cross shows the mass uncertainties which we assume to be 0.3 dex. The different panels show the relation of NSC and GC system properties with the stellar mass of the host galaxy. We note that $M_\text{GC, min}$ and $M_\text{GC, max}$ are only available for Fornax and Virgo. The few outliers in the $M_\text{gal}$ - $M_\text{GC,\,max}$ plane might be UCDs.}
    \label{fig:sample}
\end{figure*}

Our sample consists of galaxies from the Virgo and Fornax galaxy clusters, complemented by additional nucleated galaxies mainly in the Local Volume. The sample is illustrated in Fig. \ref{fig:sample}.

\subsection{ACSVCS and ACSFCS}
The Advanced Camera for Surveys (ACS) Fornax Cluster Survey (ACSFCS, \citealt{Jordan2007}) and ACS Virgo Cluster Survey (ACSVCS, \citealt{Cote2004}) are two surveys based on \textit{Hubble Space Telescope} (HST) ACS data of galaxies with  $M_\text{gal} >$ 10$^8 M_\sun$ in the Fornax and Virgo cluster, respectively. The NSCs are described in \cite{Turner2012} (Fornax) and \cite{Cote2006} (Virgo), including their sizes, $g$- and $z$-band magnitudes, and colours. For every ACSFCS and ACSVCS galaxy, photometric GC catalogues were collected \citep{Jordan2009, Jordan2015}.
These give sizes as well as $g$- and $z$-band magnitudes of photometric GC candidates in the ACSFCS and ACSVCS galaxies.

\subsubsection{Photometric masses}
Using the $g$- and $z$-band magnitudes of the GCs, we inferred their metallicities using the colour-metallicity relation (CZR) of \cite{Fahrion2020c} which was derived using a sample of spectroscopic GCs from the Fornax3D project \citep{F3D_Survey}. From the metallicities and an assumed old age of 12 Gyr, we estimated the stellar masses using the photometric predictions of the E-MILES stellar population synthesis models that give the mass-to-light ratio in the ACS $g$ band \citep{Vazdekis2016}. In this way, we derived the masses of the least massive and most massive GC ($M_\text{GC, min}$ and $M_\text{GC, max}$) in each galaxy directly from the catalogues. For the GCs, we limited the catalogues to all candidates that have a probability of being a genuine GC $p$GC $>$ 0.50 to avoid contamination. As Fig.~\ref{fig:sample} shows, $M_\text{GC, max}$ increases with galaxy mass, but there are a few outliers that could be UCDs. In Sect.~\ref{sect:discussion} we discuss the choice of $M_\text{GC, max}$.

NSCs often are chemically more complex than GCs and can also host young populations \citep{FeldmeierKrause2015}, and thus might not follow the CZR. To determine their masses, we sampled theoretical mass-to-light ratios based on their $g$- and $z$-band magnitudes within uncertainties and allow random ages between 5 and 14 Gyr. The photometric uncertainties ($\sim$ 0.2 mag) govern the resulting mass distributions, but they still allowed us to extract a mean photometric mass within $\sim$ 0.2 dex, which is above our default conservative limit of 0.3 dex that should also consider systematic uncertainties in stellar population synthesis models.

\subsubsection{Total globular system mass}
We determined the total mass of the GCS ($M_\text{GCS}$) based on the GC luminosity functions given in \cite{Villegas2010}. This work gives the mean and standard deviations in the $g$ and $z$ band of the GC luminosity distributions. Using the completeness-corrected total number of GCs $N_\text{GCs}$ (from \citealt{Peng2008, Liu2019}), we created a mock GCS that is populated by drawing 1000 times $N_\text{GCs}$ GCs from the distribution. A fraction $f_\text{red}$ (from \citealt{Peng2008, Liu2019}) of them are assigned to be red GCs, and the remaining ones are labelled as blue GCs. The total mass was then calculated by assigning a mean mass-to-light ratio to the red and blue GCs, respectively, based on the CZR.
We only differentiated into red and blue GCs for this mass calculation and do not consider a fraction which might be accreted during the assembly of the host as the colour of a GC is not sufficient to determine its origin \citep{Fahrion2020c}. Alternatively, the individual GC candidates from the photometric catalogues can be summed up, but these are not completeness-corrected and can result in lower $M_\text{GCS}$ by \mbox{$\sim$ 0.2 dex}.  The completeness correction accounts for magnitude limits within the surveys and spatial incompleteness of the ACS field-of-view.

\subsection{Other galaxies}
In addition to the ACSFCS and ACSVCS, we used a heterogeneous sample of 33 nucleated galaxies mainly in the Local Volume (D $<$ 15 Mpc), but five of those are also at larger distances. These have NSC properties presented in \cite{Georgiev2016}, \cite{Pechetti2019}, and \cite{Neumayer2020}, and the total GC system masses are from \cite{Harris2013}, \cite{Forbes2018}, and \cite{Spitler2008}.

Fig. \ref{fig:sample} illustrates that this composite sample differs from the ACSVCS and ACSFCS galaxies. At the same galaxy mass, this sample has lower NSC masses. This is caused, in part, by the larger number of LTGs in the local sample compared to Virgo and Fornax which are dominated by massive ETGs. LTG galaxies are known to follow a shallower relation between NSC and host mass \citep{Georgiev2016}. However, the two most massive galaxies in the heterogeneous sample are classified as ETGs with NSCs that are less massive than the galaxy mass would suggest. One of those ETGs is NGC\,4552 (M\,89 or VCC\,1632), an ETG in the Virgo cluster, in which \cite{Cote2004} did not report a stellar nucleus due to the presence of dust, but \cite{Lauer2005} reported a NSC with $M_\text{NSC} = 9 \times 10^7 M_\sun$. NGC\,4552 is also known to host a supermassive black hole (SMBH) with $M_\text{BH} = 5 \times 10^8 M_\sun$ \citep{Gebhardt2003}. The second massive ETG is IC\,1459 with an even more massive SMBH of $M_\text{BH} = 2 \times 10^9 M_\sun$ \citep{Saglia2016}.
The presence of these massive SMBHs could explain the lower NSC masses as SMBHs can inhibit NSC growth \citep{Antonini2013}.

For this composite sample, masses of individual GCs are not available and hence we could not use empirical estimates of the most massive and least massive observed GC for $M_\text{GC, lim}$ and $M_\text{diss}$, respectively. Therefore, we used the \citetalias{Leaman2021} prediction for $M_\text{GC, lim}$ based on the host galaxy structure. We approximated $M_\text{diss}$ for these galaxies by fitting the $M_\text{GC, min}$ - $M_\text{gal}$ relation from the ACSFCS and ACSVCS with a log-linear function and by assigning the Local Volume galaxies a randomly drawn value within this relation and its $1 \sigma$ scatter. In the ACS galaxies, the mass of $M_\text{GC, min}$ seems to decrease with increasing galaxy mass (see bottom right panel in Fig. \ref{fig:sample}). This could be a sampling effect due to the larger number of GCs in the more massive galaxies causing the GC mass function to be better sampled. In general, $M_\text{GC, min}$ only varies within $\sim$ 0.4 dex from the lowest to the highest galaxy masses.

\begin{figure*}
    \centering
    \includegraphics[width=0.99\textwidth]{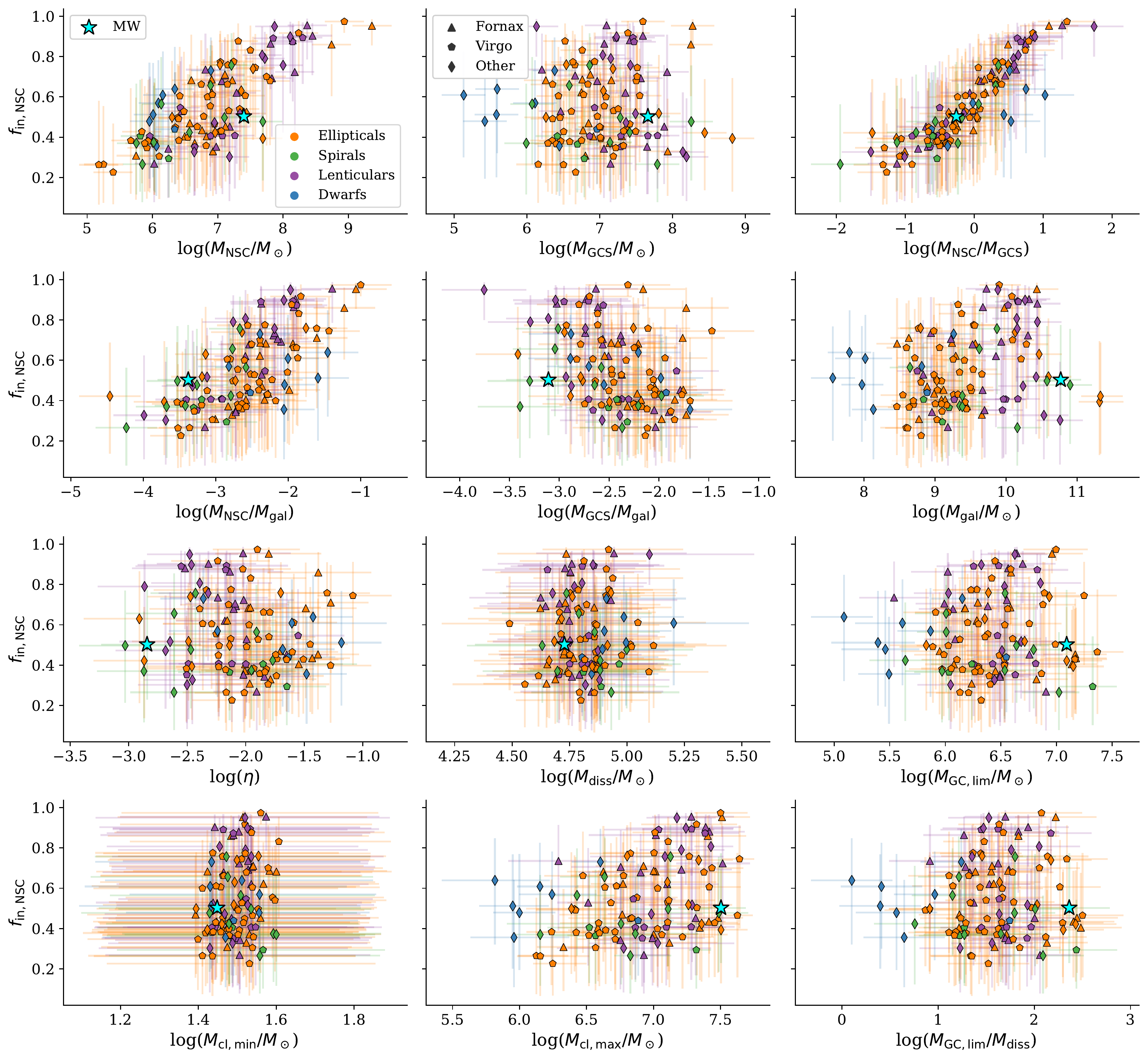}
    \caption{NSC mass fraction formed from in situ star formation ($f_\text{in, NSC}$) versus model parameters. The model was fitted to $M_\text{NSC}$ and $M_\text{GCS}$ (first row), while the other parameters were introduced as priors in the model. We used the same colours and symbols as in Fig. \ref{fig:sample}. The error bars refer to the 16th and 84th percentile of the distribution (see Fig. \ref{fig:FCC47_corner}).}
    \label{fig:results}
\end{figure*}

\section{Results}
\label{sect:results}
We applied the model predictions described in Sect. \ref{sect:model} to measurements of $M_\text{NSC}$ and $M_\text{GCS}$ of 119 nucleated galaxies to infer an estimate of $f_\text{in, NSC}$. Figure \ref{fig:results} shows the resulting relations between $f_\text{in, NSC}$ and the various input parameters. The error bars refer to the 16th and 84th percentile of the posterior distributions (see Fig. \ref{fig:FCC47_corner}).

\subsection{Trend with NSC and GCS masses}
As the upper left corner of Fig. \ref{fig:results} illustrates, the marginalised values of $f_\text{in, NSC}$ show an increase with the NSC mass $M_\text{NSC}$. Our model predicts that low mass NSCs are formed predominantly via the accretion of GCs ($f_\text{in, NSC} \lesssim 0.50$), while the most massive NSCs (those with NSCs approximately $\sim 1 - 10\%$ of the host mass) reach $f_\text{in, NSC} \gtrsim 0.90$. We note that $f_\text{in, NSC}$ shows no significant relation with $M_\text{GCS}$ or $M_\text{gal}$, suggesting that $f_\text{in, NSC}$ is not strongly affected by potential incompleteness or contamination in the GC population.

The tightest correlation is seen between $f_\text{in, NSC}$ and the ratio between the NSC and GCS mass, log$(M_\text{NSC}/M_\text{GCS})$. Where both have comparable masses, the model predicts roughly equal contributions from both formation channels. This is likely a direct result of a single model threshold mass $M_\text{GC, lim}$ sub-dividing a parent power-law distribution of GC masses.
In systems where the NSC is significantly more massive than the GCS, high in situ fractions are required to simultaneously reproduce the observed GC and NSC masses. At the same time, NSCs, which only constitute a fraction of the GCS, are found to have predominantly formed from the accretion of GCs.  We show in Sect. \ref{sect:galaxy_accretion} that even significant amounts of galactic scale accretion are likely not enough to drop the in situ fractions for the NSCs in these systems significantly. The trend with the NSC-to-GCS mass ratio likely also reflects the behaviour of $f_\text{in, NSC}$ with the GCS-to-host mass ratio. Galaxies with more massive GC systems relative to the stellar mass of the host favour low in situ fractions as those are likely the systems where the GCSs dominate the NSC-to-GCS mass ratio.

For convenience, we fitted the trends of $f_\text{in, NSC}$ with  $(M_\text{NSC}/M_\text{GCS})$ and $M_\text{NSC}$ with a function of the following form:
\begin{equation}
    f_\text{in, NSC} = \beta\,\text{tanh}(x - \alpha) + (1 - \beta),
    \label{eq:fitting_funct_2}
\end{equation}
with $x$ = log$(M_\text{NSC}/M_\text{GCS})$ or $x$ = log($M_\text{NSC}/M_\sun)$. We note that $\alpha$ describes the saddle point and $\beta$ provides an offset at small $x$ and ensures that $f_\text{in, NSC} \leq 1$. The fits were performed with \textsc{emcee} using symmetrised uncertainties of $f_\text{in, NSC}$, but a least-squared fit gives comparable results.

\begin{figure*}
    \centering
    \includegraphics[width=\textwidth]{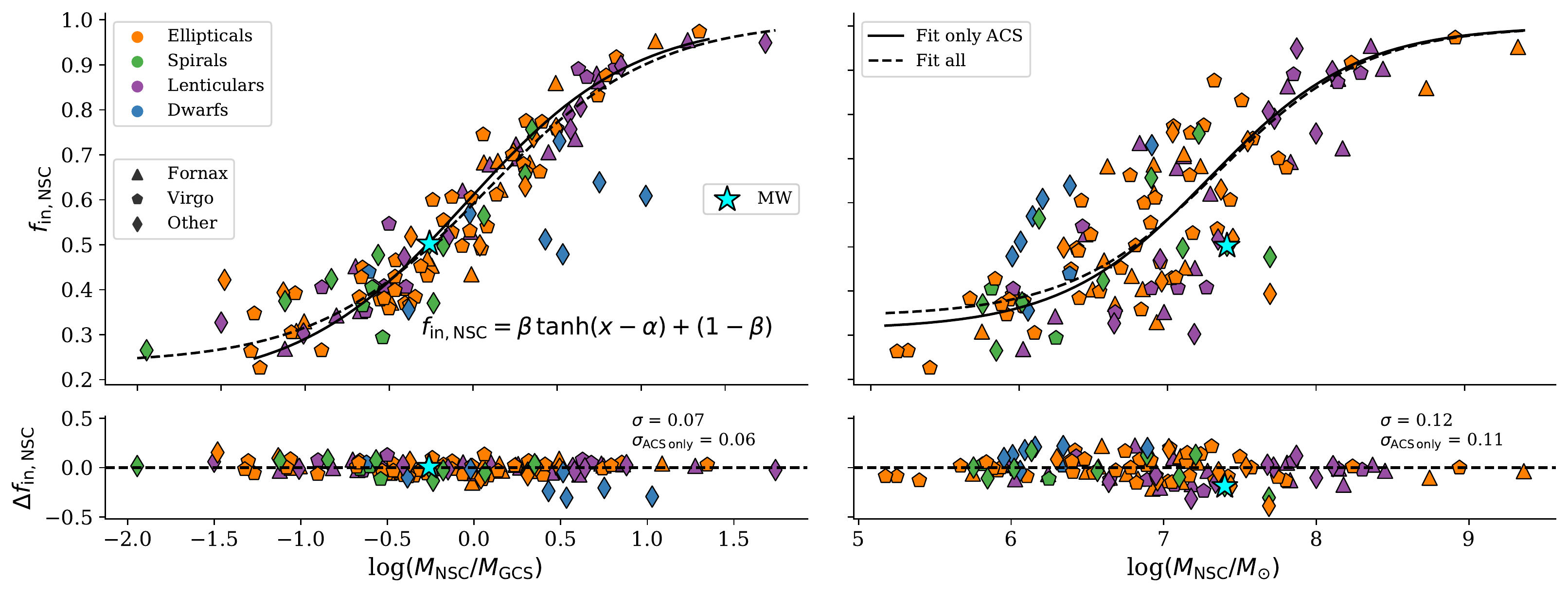}
    \caption{
    In situ mass fraction $f_\text{in, NSC}$ versus log$(M_\text{NSC}/M_\text{GCS})$ (\textit{left}) and $M_\text{NSC}$ (\textit{right}) as seen in Fig. \ref{fig:results}. The solid line shows the fit using only the ACS data (Fornax and Virgo galaxies), and the dashed line when using all data. The bottom panels show the residual scatter. The standard deviations are shown in the label.}
    \label{fig:fitting_funct}
\end{figure*}

Figure \ref{fig:fitting_funct} shows the fits and the residual scatter. We indicate the standard deviation of the residual considering all the data and when only including the ACSVCS and ACSFCS data. As mentioned above, the tightest relation is seen for $f_\text{in, NSC}$ versus  log$(M_\text{NSC}/M_\text{GCS})$, but only for the ACS data. Including data from the Local Volume increases the scatter. The best-fitting values are reported in Table \ref{tab:fitting_function_results}. These show that the switch from GC-accretion dominated formation to in situ formation happens at a characteristic NSC mass of log($M_\text{NSC}/M_\sun) \sim 7.1$.

Figures \ref{fig:results} and \ref{fig:fitting_funct} show that the relation between $f_\text{in, NSC}$ and log$(M_\text{NSC}/M_\text{GCS})$ is clearly tightest for the ACS galaxies, but some other galaxies show deviations from the relation. These are in particular the few irregular dwarf and spheroidal galaxies in the Local Volume \citep{Georgiev2009a, Georgiev2009} which are characterised by large values of log$(M_\text{NSC}/M_\text{GCS})$ at low galaxy masses. As Fig. \ref{fig:sample} shows, the NSC in these galaxies makes up a significant fraction of the total galaxy mass and of the total mass in star clusters. They may have had a more efficient bound cluster formation, increasing the typical spread in the ratio of $M_\text{NSC}/M_\text{GCS}$. In these galaxies, already the in spiral of a single GC significantly increases the value of log$(M_\text{NSC}/M_\text{GCS})$ (but not $M_\text{NSC}$ so much) due to the low number of GCs in total. This shot noise could explain why they are outliers in the $f_\text{in, NSC}$ versus log$(M_\text{NSC}/M_\text{GCS})$ relation, but not in $f_\text{in, NSC}$ versus log$(M_\text{NSC})$.
Hence these are examples of the stochastic nature of how this ratio will evolve in the case of dwarf galaxies where a few clusters may form in total.

On the other hand, the two ETGs NGC\,4552 and IC\,1459 discussed above now stand out for having larger $f_\text{in, NSC}$ values than other galaxies with  log$(M_\text{NSC}/M_\text{GCS})$ between $-1.5$ and $-1.0$. These two galaxies are the two most massive ones in the sample and as discussed above, they are known to host SMBHs which could have altered the NSC mass evolution. Galaxy size or cluster formation efficiency may also play a role in the location of galaxies on this diagram. However, these two galaxies are not outliers in the $f_\text{in, NSC}$ - log($M_\text{NSC}/M_\odot$) plane.

\begin{table*}[h]
    \centering
    \caption{Fitting function results, see Eq. \ref{eq:fitting_funct_2}.}
    \begin{tabular}{c c c c c c }
     x & Data & $\alpha$ & $\beta$ &  $\sigma_\text{res}$ & valid range \\ \hline \hline
     log($M_\text{NSC}/M_\text{GCS})$ & ACS & $-0.04^{+0.14}_{-0.15}$ & $0.40^{+0.05}_{-0.04}$ & 0.06 & $-1.3 < \text{log}(M_\text{NSC}/M_\text{GCS}) < 1.4$\\
     log($M_\text{NSC}/M_\text{GCS}$) & all &
     $0.07^{+0.11}_{-0.13}$ & $0.38^{+0.04}_{-0.03}$ &
      0.07 & $-2.0 < \text{log}(M_\text{NSC}/M_\text{GCS}) < -1.8$\\
     log($M_\text{NSC}/M_\sun$) & ACS &  $7.34^{+0.15}_{-0.14}$ & $0.33^{+0.02}_{-0.02}$ & 0.11 & $5.1 < \text{log}(M_\text{NSC}/M_\sun) < 9.4$\\
    log($M_\text{NSC}$/$M_\sun$) & all & $7.28^{+0.19}_{-0.18}$ & $0.34^{+0.03}_{-0.03}$  & 0.12 & $5.1 < \text{log}(M_\text{NSC}/M_\sun) < 9.4$ \\ \hline
    \end{tabular}
    \label{tab:fitting_function_results}
    \tablefoot{Uncertainties on $\alpha$ and $\beta$ from MCMC sampling; $\sigma_\text{res}$ refers to the standard deviation of the residual scatter.}
\end{table*}

\subsection{Trends with model parameters}
Importantly, the normalisation factor $\eta$, describing how much mass forms in clusters, does not correlate with the in situ fraction.  This is perhaps expected as the reproduction of the NSC and GCS masses is influenced by the mass sub-divisions of the power law mass function, rather than its integrated value. The most massive cluster that ever formed, $M_\text{cl, max}$, shows a positive correlation with $f_\text{in, NSC}$ and reaches the upper limit of the prior for galaxies with high in situ fractions. Studies of molecular cloud formation and statistical arguments on GC populations \citep{ReinaCampos17,Norris19} suggest that this value is unlikely to significantly exceed the range of our default prior of $7.6 \times 10^7 M_\sun$; however, we discuss further how the choice of this prior affects the results in Sect. \ref{sect:model_setup}.

We cannot identify a significant environmental dependence on any of the trends of $f_\text{in, NSC}$ or on the galactic or GC system properties as shown in Fig. \ref{fig:results}. There may be a hint that predominantly the field galaxies of the composite sample show lower values of $f_\text{in, NSC}$ at a fixed GC-host mass ratio ($M_\text{GCS}/M_\text{gal}$); however, within the uncertainties, it is consistent with the Fornax and Virgo samples. Further, the composite sample shows larger scatter in the $f_\text{in, NSC}$ - ($M_\text{NSC}/M_\text{GCS}$) phase space. This is likely caused by the shallower $M_\text{NSC}$ - $M_\text{gal}$ relation observed for this sample (Fig. \ref{fig:sample}), while the $M_\text{GCS}$ - $M_\text{gal}$ relation agrees with the ACSFCS and ACSVCS sample.

The lack of environmental dependence for $f_\text{in, NSC}$ may indicate that the processes (GC in spiral and in situ star formation) are both occurring in the inner ($< R_\text{eff}$) regions of the galaxy and are not significantly influenced by {minor} tidal effects affecting the outskirts. However, it also indicates that the observables which drive the model ($M_\text{NSC}, M_\text{GCS},    M_\text{GC, lim}$) have also not been significantly altered by environmental processes at a fixed galaxy density.

\begin{table*}[h]
    \centering
    \caption{Description of model sub-sets.}
    \begin{tabular}{l|l}
      Model Run   & Description  \\ \hline \hline
    Default & Default approach described in Section 2 \\
    2nd brightest GC & Adopt the 2nd brightest GC for $M_\text{GC, lim}$ \\
    Model $M_\text{GC, lim}$ & Adopt the \citetalias{Leaman2021} prediction for $M_\text{GC, lim}$ \\
    $M_\text{cl, max} < 10^{8.8}$ & Upper limit for $M_\text{cl, max}$ set to $10^{8.8} M_\sun$\\
    Variable accretion & Allow for a fraction $f_\text{acc, gal}$ of GCs to be accreted \\
    Adaptive priors & Allow for variable accretion, and $f_\text{in, NSC}$ constrained by adaptive priors (see \citetalias{Leaman2021}) \\
    Simulation $f_\text{acc, gal}$ & Adopt $f_\text{acc}$ as a function of $M_\text{gal}$ from the EAGLE simulations \citep{Davison2020} \\ \hline
    \end{tabular}
    \label{tab:versions}
\end{table*}

\section{Discussion}
By applying an analytic model for NSC formation to observations of the NSC and GC system masses in 119 galaxies, we show in Sect. \ref{sect:results} how we estimated the relative contribution of in situ star formation to NSC growth. The predicted $f_\text{in, NSC}$ shows a strong correlation with the mass ratio of a galaxy's NSC to GC system mass; however, given the simplicity of the model, we wish to understand whether additional factors may bias this result.

\label{sect:discussion}

\subsection{Dependence on parameter setups}
\label{sect:model_setup}
\begin{figure}
\centering
\includegraphics[width=0.49\textwidth]{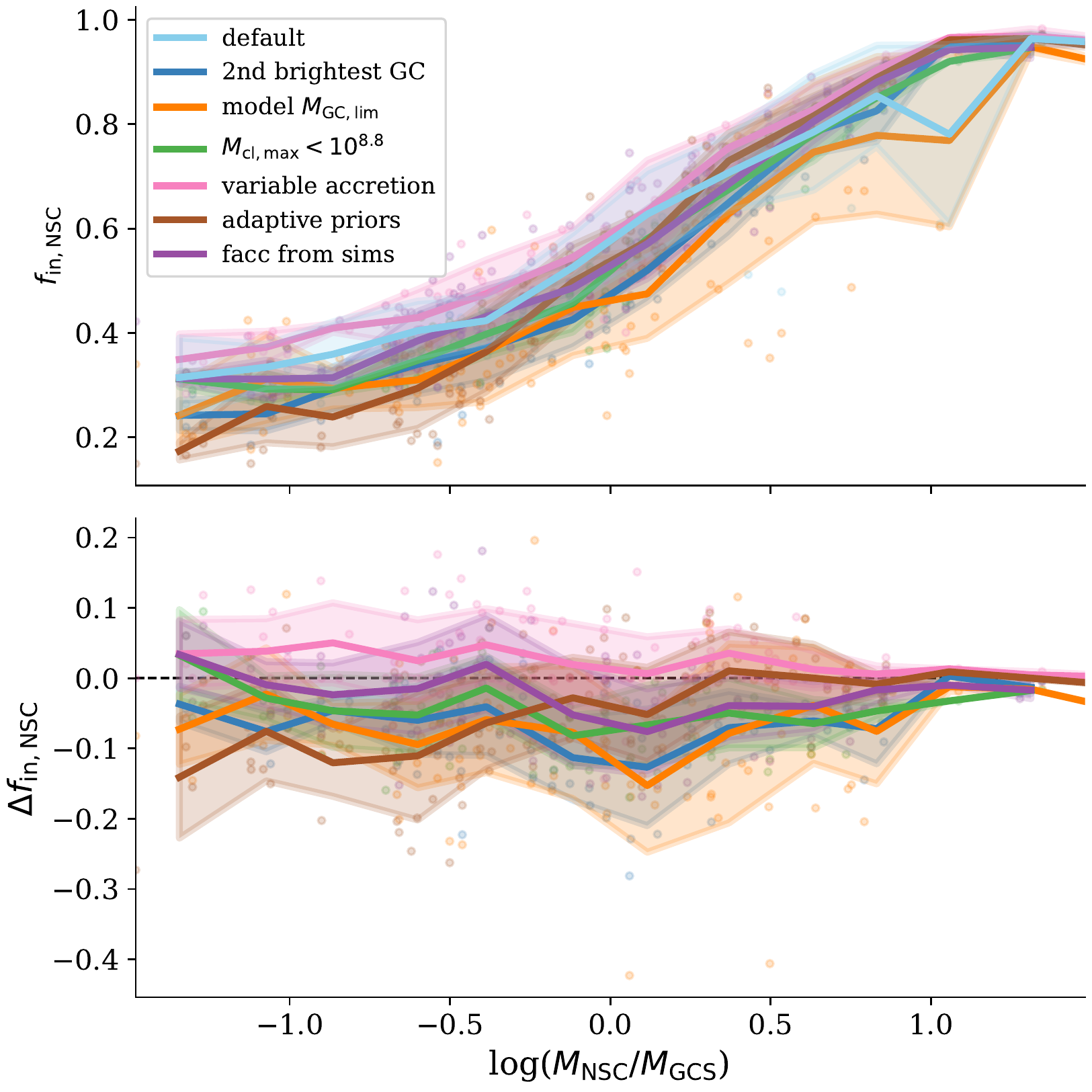}
\caption{Influence of model settings on the predicted in situ NSC mass fractions. \textit{Top}: $f_\text{in, NSC}$ in relation with the ratio of NSC and GCS mass for the different settings. Coloured lines show median-binned data with 1$\sigma$ scatter for visualisation. \textit{Bottom}: Differential in situ mass fraction relative to the default model. The different settings are summarised in Table \ref{tab:versions}.}
\label{fig:MGClim}
\end{figure}

To explore the influence of assumptions within the model and assumed priors, we ran the model on the ACSFCS and ACSVCS data with different choices, as described in Table \ref{tab:versions}. The resulting changes to our key variable ($f_{\text{in, NSC}}$) and its relation to galactic star cluster system properties are shown in Fig. \ref{fig:MGClim}. Using the NSC mass shows comparable changes.

The limiting GC mass for accretion into the NSC, $M_\text{GC, lim}$, is a key parameter in our model. The most massive GC observed today should give a reasonable approximation of this limiting mass, but as shown in \citetalias{Leaman2021}, a contamination from UCDs could bias it to higher masses. Therefore, we also ran the model using two alternative estimates for this parameter: 1) using the second most massive observed GC (model run `$M_\text{GC, 2nd max}$'), and 2) deriving $M_\text{GC, lim}$ from the galaxy's size and mass as in \citetalias{Leaman2021} (`model $M_\text{GC, lim}$'). While the overall relation with $M_\text{NSC}/M_\text{GCS}$ is robust, the highest in situ fractions are generally found when using the default approach because both the model prediction of $M_\text{GC, lim}$ and $M_\text{GC, 2nd max}$ tend to be lower than $M_\text{GC, max}$ used in the default approach. This allows for more massive GCs merging into the NSC, consequently reducing $f_\text{in, NSC}$ by up to 10\%.

We next tested the influence of $M_\text{cl, max}$, the mass of the most massive GC that ever formed. The posterior values returned by the default model for $M_\text{cl,max}$ are well within the observed values seen in the Milky Way and Local Volume, and even merging systems \citep{Renaud2018}, but the upper limit to the prior of \mbox{$M_\text{cl, max} = 7.6 \times 10^7 M_\sun$} \citep{Norris2019} is reached for some of the highest $f_\text{in, NSC}$ systems. Increasing this limit by one order of magnitude to $\sim 7 \times 10^8$ (`$M_\text{cl, max} < 10^{8.8}$' model) can reduce the in situ fractions in these galaxies by $\sim$ 10 percent.  While this is expected because it allows for more massive GCs to be formed, neither this nor the changes to $M_\text{GC, lim}$ alter the trend of $f_\text{in, NSC}$ versus log($M_\text{NSC}/M_\text{GCS}$).

\subsection{Galaxy accretion}
\label{sect:galaxy_accretion}
Our model regards galaxies as isolated systems as observed today and does not consider their mass assembly via mergers or the accretion of dwarf galaxies. However, these mergers alter the GCs that are available for accretion into the NSC, the host galaxy GC system mass, and the dynamical friction conditions. To test the influence of galaxy accretion, we also tested a model setup where an additional parameter, $f_\text{acc}$, is introduced. This parameter describes the fraction of GCs and stellar mass accreted by a galaxy due to mergers, and it scales the observed GCS mass and galaxy mass via $(1 - f_\text{acc, gal}) M_\text{GCS}$ and $(1 - f_\text{acc, gal}) M_\text{gal}$ (see Appendix D in \citetalias{Leaman2021}).

We implemented the galaxy accretion fraction both with a flat prior from zero to unity (model setup `variable accretion') and with a prior that is adapted continuously via the upper and lower limits on $f_\text{in, NSC}$ as given in \citetalias{Leaman2021} (model setup `adaptive priors'):
\begin{equation}
    f_\text{in, NSC} \geq 1 - \left(\frac{\eta(1-f_\text{acc, gal})^2 M_\text{gal} M_\text{GCS}}{M_\text{NSC} M_\text{GC, lim} (1 + \text{ln} \frac{M_\text{GC, lim}}{M_\text{diss}})} \right) and
    \label{fin_lower}
\end{equation}
\begin{equation}
    f_\text{in, NSC} \leq 1 - \left(\frac{M_\text{GC, lim}M_\text{GCS}}{\eta M_\text{gal} M_\text{NSC}} \right).
    \label{fin_up}
\end{equation}

Introducing $f_\text{acc, gal}$ with flat priors leads to higher values of $f_\text{in, NSC}$ since the same NSC mass has to be reproduced with a reduced GC system. The effect is larger in galaxies that have low in situ fractions in the default model because those are the systems where the GCS was dominating previously. In contrast, when using the adaptive priors, the upper limit requires lower in situ fractions for galaxies where the $M_\text{NSC}/M_\text{GCS}$ ratio is dominated by the GCS, further reducing $f_\text{in, NSC}$.

As discussed in \citetalias{Leaman2021}, while limits on both the NSC in situ mass fraction and galaxy accretion fraction possibly depend on the particular stellar mass and galaxy size, we do not expect the model to uniquely infer both the NSC in situ mass fraction $f_\text{in, NSC}$ and the galaxy accretion fraction $f_\text{acc, gal}$.  This is reasonable as both parameters alter the mass ratio of the NSC to GC system mass in degenerate ways. For this reason, we produced a model run using trends of $f_\text{acc, gal}$ with the host galaxy mass from the EAGLE simulations of galaxy formation \citep{Davison2020} (model setup `$f_\text{acc}$ from simulations'). Since these theoretical predictions find low accretion fractions between $\sim$ 10 and 50 \% for the galaxy mass range studied here, this results in $\text{in situ}$ fractions similar to the default model.

In summary, the exact setup of our model can affect the resulting $f_\text{in, NSC}$ by $\sim 20\%$, except for the largest ratios of $M_\text{NSC}/M_\text{GCS}$. However, the general trend with NSC mass  remains unaltered by these systematics.

\subsection{Expectations from stellar populations}
\begin{figure}
\centering
\includegraphics[width=0.49\textwidth]{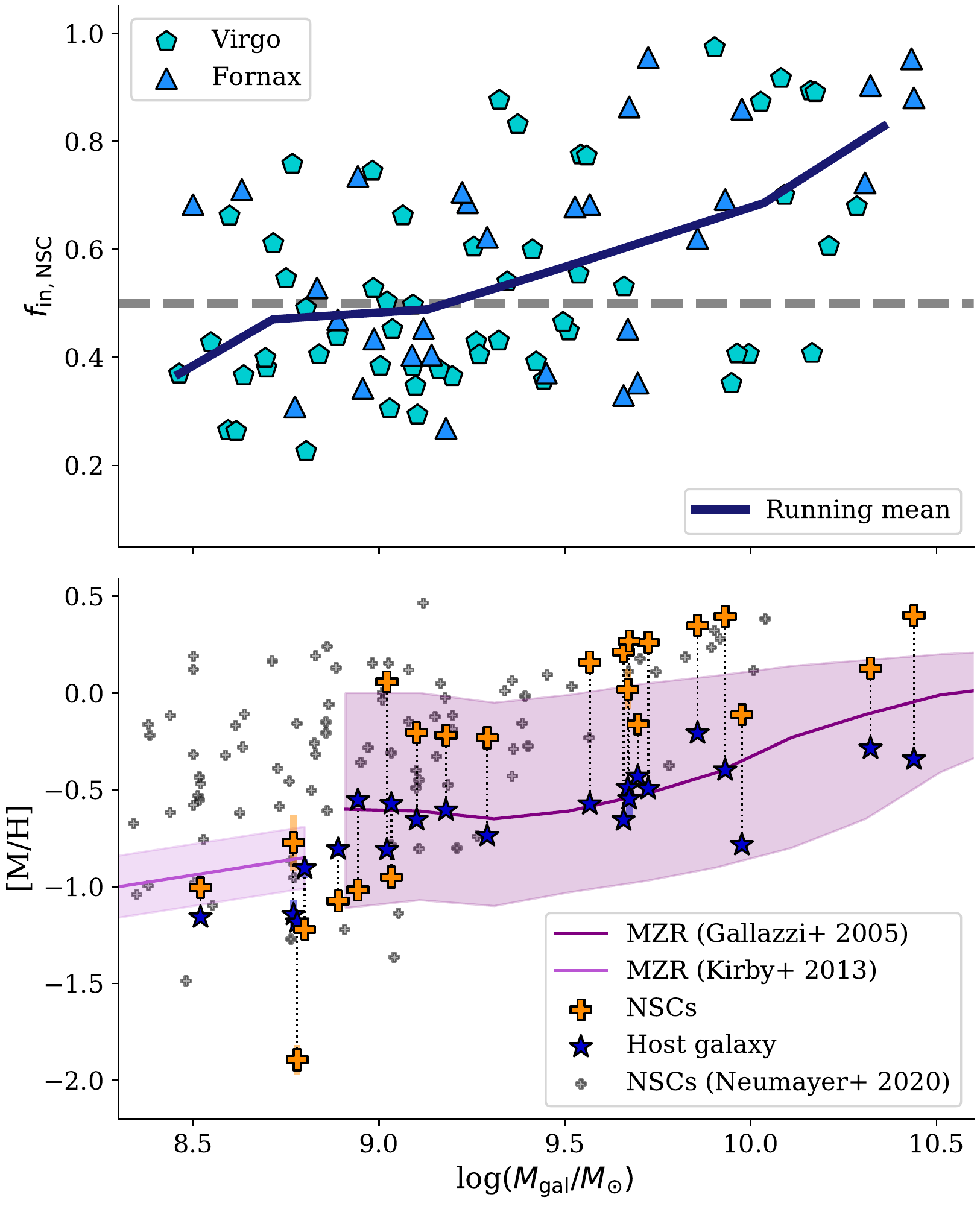}
\caption[\textit{Top}: in situ mass fraction $f_\text{in, NSC}$ versus galaxy mass for ACS galaxies.]{In situ mass fraction $f_\text{in, NSC}$ versus galaxy mass for galaxies in Virgo and Fornax. The red line is the moving average (not considering the uncertainties in $f_\text{in, NSC}$) that crosses $f_\text{in, NSC}$ = 50\% for galaxy masses $M_\text{gal} \sim  1.5 \times 10^{9} M_\sun$. \textit{Bottom}: Mass-metallicity relation (MZR) of galaxies \citep{Gallazzi2005, Kirby2013}, with NSCs and host galaxy metallicities plotted in the coloured plus and star symbols from the measurements presented in \cite{Fahrion2021}. Small grey plus symbols refer to other literature NSC metallicities as compiled in \cite{Neumayer2020}.}
\label{fig:model_f_in_vs_M_gal}
\end{figure}

As \cite{Neumayer2020} argue, insights from stellar population analysis of NSCs indicate that at low galaxy masses, GC accretion constitutes the dominant NSC formation channel, whereas in situ formation might become more influential for galaxies more massive than \mbox{$M_\ast \sim 10^9 M_\sun$.}
This argument is mainly based on the observation that NSCs in lower mass galaxies tend to be more metal-poor than an {average} galaxy of that mass from empirial mass-metallicity relations (MZRs, bottem panel of Fig. \ref{fig:model_f_in_vs_M_gal}), while NSCs in massive galaxies often show high metallicities, young subpopulations, flattened structures, and can be kinematically complex \citep{Seth2008, Paudel2011, Spengler2017, Lyubenova2013, Lyubenova2019, Fahrion2019b}.

In \cite{Fahrion2021}, we could confirm such a trend more directly with the stellar population analysis of NSCs and their individual hosts in 25 ETGs and dwarf ellipticals. This showed that the dominant NSC formation channel transitions from GC accretion to in situ star formation with an increasing galaxy and NSC mass. We found that the transition happens at masses at $M_\text{gal} \sim 10^9 M_\sun$ and $M_\text{NSC} \sim 10^7 M_\sun$, respectively.

Using the semi-analytical model presented here, we quantified the fraction of the NSC mass assembled via in situ star formation, $f_\text{in, NSC}$, and we show that it is strongly correlated with the total NSC mass (see Fig. \ref{fig:fitting_funct}). The model can reproduce the transition at NSC masses $M_\text{NSC} \sim 10^{7} M_\sun$, but the correlation with galaxy mass is weaker as Fig. \ref{fig:model_f_in_vs_M_gal} shows. This figure again shows the in situ NSC fraction versus galaxy mass, but now only focusing on the homogeneous ACS sample of ETGs in Virgo and Fornax. The moving average reaches $f_\text{in, NSC}$ = 50\% for galaxy masses $M_\text{gal} \sim  1.5 \times 10^{9} M_\sun$, but the scatter is significant. A trend with host mass in the model is a secondary effect as more massive galaxies tend to have more massive NSCs (see Fig. \ref{fig:sample} or \citealt{Ordenes2018, SanchezJanssen2019}), but at the same time have a large scatter in the ratio of NSC-to-GC system mass (Fig. \ref{fig:sample}), the most sensitive parameter in our model. Due to the limited number of galaxies for which a detailed stellar population analysis could identify the dominant NSC formation channel \citep{Fahrion2020a, Fahrion2021}, it is still unclear how large the scatter in the dominant NSC formation channel is found to be from observations.

The Milky Way NSC is close enough to study individual stars, although in the infrared due to strong galactic extinction. It is known to be chemically and kinematically complex, containing multiple stellar populations with varying metallicities, stellar ages, and orbital configurations (e.g. \citealt{Lu2013, Feldmeier2014, Feldmeier2017, Do2018, Feldmeier2020}). Our model finds that roughly half of the MW NSC mass was formed via in situ star formation with $f_\text{in, NSC, MW} = 0.54^{+0.33}_{-0.25}$. Observationally, there is also evidence that both NSC formation channels have contributed to its growth. The presence of young stars indicates some level of in situ star formation, although the rotation in the NSC can also be explained from the merger of multiple GCs \citep{Tsatsi2017}.
Recently, a minor metal-poor component was discovered that constitutes $\sim 7 \%$ of the stars \citep{Feldmeier2020} and might have formed from an infalling GC or even a dwarf galaxy \citep{ArcaSedda2020, Do2020}.

\begin{figure*}
    \centering
    \includegraphics[width=0.96\textwidth]{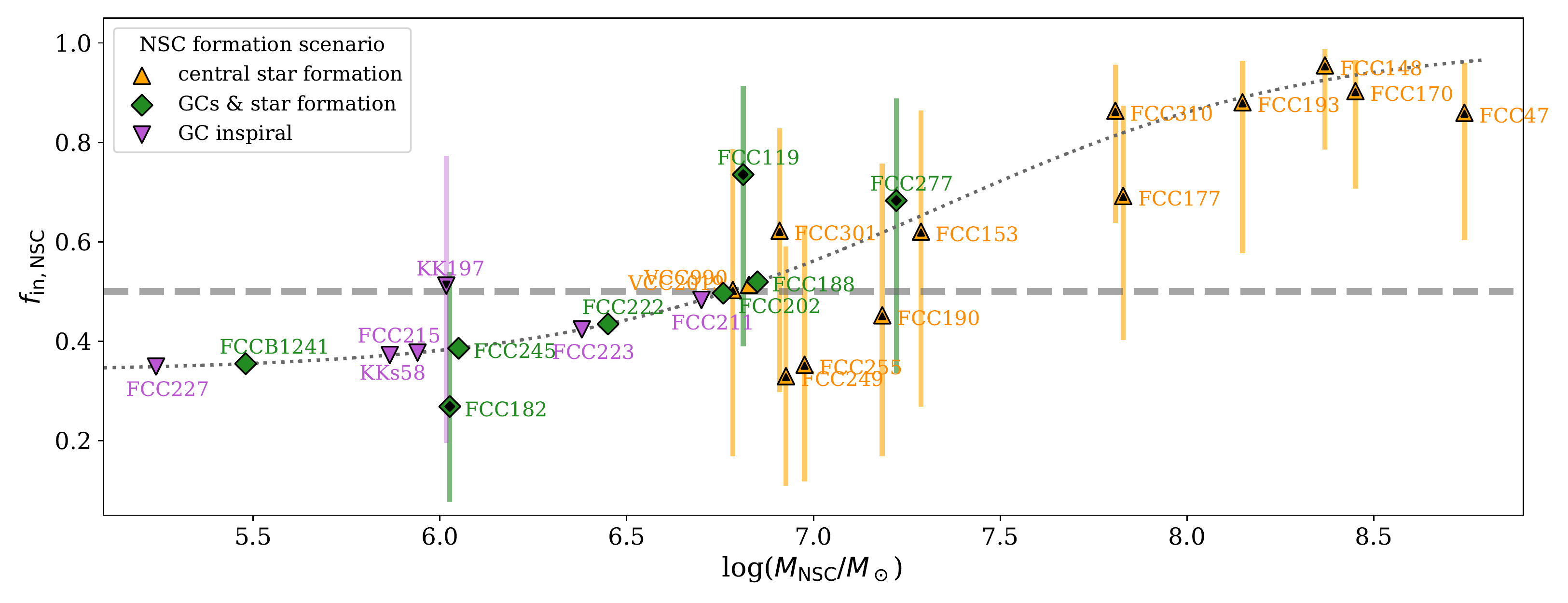}
    \caption[Comparison between modelling and stellar population results.]{NSC mass versus in situ NSC mass fraction for the 25 galaxies studied in \cite{Fahrion2021} as well as the two dwarf galaxies, KK\,197 and KKs\,58 \citep{Fahrion2020a}. The symbols indicate the dominant NSC formation channel as identified in \cite{Fahrion2021}, based on stellar population analysis of the NSCs in comparison to their host galaxies. Galaxies that were modelled in this work are highlighted with a small black symbol and show the inferred uncertainties on $f_\text{in, NSC}$. For the other galaxies, the fitting function for the ACS galaxies as in Table \ref{tab:fitting_function_results} was used (dotted line). The grey dashed line shows $f_\text{in, NSC}$ = 50\%.}
    \label{fig:model_comparison_to_pops}
\end{figure*}

For an important independent test of our model, we compare the results from our semi-analytical model to the stellar population analysis of the 25 ETGs presented in \cite{Fahrion2021} and the two dwarfs KK\,197 and KKs\,58 \citep{Fahrion2020a} in Fig. \ref{fig:model_comparison_to_pops}. This figure shows the inferred in situ mass fraction either from the model directly or from the fitting function of NSC mass versus $f_\text{in, NSC}$ (Table \ref{tab:fitting_function_results}) for those galaxies that do not have information on their GC system available. The different symbols indicate the dominant NSC formation channel as was identified from stellar population analysis (see fig. 6 in \citealt{Fahrion2021}).

This figure illustrates that our semi-analytical model and this completely independent stellar population analysis from spectroscopy agree in their determination of the dominant NSC formation channel for the least massive and most massive NSCs. This is an important test of the purely analytical model presented here, and further observations and analysis similar to the ones presented in \cite{Fahrion2021} will be useful in the future to improve the interpretations and this model.

For the highest NSC masses (log($M_\text{NSC}) > {7.5}$), the model reports high values of $f_\text{in, NSC}$, which is in agreement with the result from the stellar population analysis that identified the in situ channel as the dominant NSC formation pathway in these galaxies, as argued based on the metal-rich populations in the NSC and often complex star formation histories \citep{Fahrion2021}. In addition, for the low-mass NSCs (log($M_\text{NSC}) < {6.5}$), the model predicts low in situ fractions, which is in agreement with stellar population analysis that identified those NSCs to form at least part of their mass from accreted GCs as evident from their old, metal-poor populations. These are systems where the NSC is significantly more metal-poor than the host.

At intermediate NSC masses around log($M_\text{NSC}) \sim {7}$, the model shows scatter in the inferred in situ NSC fraction and found low values of $f_\text{in, NSC} < 0.4$ for three galaxies (FCC\,249, FCC\,255, and FCC\,190) in which the stellar population analysis identified in situ star formation to be the dominant NSC formation channel \citep{Fahrion2021}. It is likely that the model prefers low values of $f_\text{in, NSC} < 0.4$ for these galaxies because they have rather low NSC masses for their galaxy mass ($M_\text{gal} \sim 5 \times 10^9 M_\sun$), but high NSC-to-GC system mass ratios.

It is perhaps not surprising that systems that our model indicate to have close to an equal contribution of in situ and GC accretion would have a large diversity of possible chemical evolution signatures. The comparison of the qualitative stellar population characterisation with the numerical in situ fraction predicted by this model will undoubtedly be complex due to the diversity in stellar populations and the purely dynamical basis of the current model. For example, plausible NSC growth could occur through metal-rich GCs forming near the centre in a nuclear ring, or conversely low metallicity gas inflow, which would complicate the interpretation of the stellar populations in terms of a single formation channel. While the overall trends are reproduced in the model, the behaviour of systems with intermediate NSC and galaxy masses can deviate from the model predictions. In a future development of the purely mass-based model presented here, we plan to include predictions for stellar population parameters to better explore the evolutionary history of individual NSC-galaxy systems.

\section{Concluding remarks}
\label{sect:conclusion}

In this paper, we have employed a semi-analytic model of NSC formation \citep{Leaman2021} to determine the dominant NSC-formation channel (either the in situ star formation or the accretion of GCs) in a sample of 119 nearby galaxies in the Local Volume, the Fornax and the Virgo  clusters, with galaxy masses from $3 \times 10^{7}$ to $3 \times 10^{11} M_\odot$, whose NSC and GC system masses have been previously determined. We quantified the fraction of the NSC mass assembled via in situ formation, $f_\text{in, NSC}$, in each NSC and explored its dependence  on various properties of the host galaxy and its GC system.

Our analysis revealed that $f_\text{in, NSC}$ strongly correlates with the NSC mass, such that low-mass NSCs have low in situ mass fractions, meaning that they were predominantly formed via the merging of GCs. Higher-mass NSCs have high $f_\text{in, NSC}$ indicating that they  mostly assembled through in situ star formation. The transition between the relative importance of the GC-accretion and  in situ star formation happens for NSC masses $\sim 10^7 M_\sun$.  In galaxies that host the most massive NSCs  (those that encompass $\sim10\%$ of the host galaxy mass), we find that the in situ channel is the dominant one, which is responsible for the build-up of more than $\sim 90\%$ of the NSC mass. As more massive NSCs are found in more massive galaxies, this further indicates a shift from GC-accretion dominated formation to in situ dominated formation with galaxy mass, as suggested by recent studies relying on stellar population properties \citep{Neumayer2020, Johnston2020, Fahrion2020a, Fahrion2021}.

We further found a stronger correlation between $f_\text{in, NSC}$ and the ratio of the NSC-to-host GCS mass. The GC-accretion channel is dominant in galaxies in which the GC system is more massive than the NSCs. Both NSC formation channels equally contribute in galaxies with comparable masses of the NSC and the GCS. While this is a purely analytic model, we importantly find that completely independent estimates of the dominant NSC formation channel from independent spectroscopic stellar population analysis is in agreement with our trend of the in situ fraction with NSC mass.

The semi-analytic model presented in this work solely relies on dynamical arguments and only exploits structural properties of the host galaxy, NSC and GCS. In a future work, we will incorporate stellar population predictions to allow for a more detailed comparison to spectroscopic observations.

\begin{acknowledgements}
We thank the anonymous referee for insightful comments that have helped to improve this manuscript. KF acknowledges support from the European Space Agency (ESA) as an ESA Research Fellow. GvdV acknowledges funding from the European Research Council (ERC) under the European Union's Horizon 2020 research and innovation programme under grant agreement No 724857 (Consolidator Grant ArcheoDyn).
\end{acknowledgements}

\bibliographystyle{aa} 
\bibliography{References}
\end{document}